\documentclass[
reprint,
amsmath,amssymb,
aps,
prb,
nobibnotes,
]{revtex4-2}

\usepackage{graphicx}
\usepackage{bm,bbm}
\usepackage{url,hyperref}
\usepackage{siunitx}
\usepackage{tikz} 

\everymath{\displaystyle}
{
\renewcommand*{\bm}[1]{#1}
}
\newcommand{\sgn}{\operatorname{sgn}} 
\newcommand{\Tr}{\operatorname{Tr}} 
\renewcommand{\Re}{\operatorname{Re}} 
\renewcommand{\Im}{\operatorname{Im}} 
\renewcommand{\vec}[1]{\bm{#1}} 
\renewcommand{\d}[0]{\mathrm{d}} 

\begin{document}

\title{Analytical formulas for far-field radiated energy and angular momentum of metallic thin films}

\author{Hankun Zhang} 
\email{hankun_z@u.nus.edu}
\affiliation{Department of Physics, 
National University of Singapore, 
Singapore 117551, 
Singapore}

\author{Yuhua Ren} 
\affiliation{Department of Physics, 
National University of Singapore, 
Singapore 117551, 
Singapore}

\author{Ho-Yuan Huang} 
\affiliation{Department of Physics, 
National University of Singapore, 
Singapore 117551, 
Singapore}

\author{Jian-Sheng Wang}
\email{phywjs@nus.edu.sg}
\affiliation{Department of Physics, 
National University of Singapore, 
Singapore 117551, 
Singapore}

\date{\today}

\begin{abstract}
We investigate far-field radiation of energy, linear momentum, and angular momentum from two-dimensional electron systems, focusing on metallic thin films described by the Drude conductivity. 
Using the Keldysh formalism within the non‑equilibrium Green’s function framework, we derive analytical expressions for radiative power, force, and torque. 
To enable angular momentum radiation, an out-of-plane magnetic field is applied to break reciprocity, resulting in gyrotropic terms in the permittivity tensor.
By approximating the emitter as a thin film, the photon Green's functions can be solved analytically. 
Expressions for the Poynting vector and Maxwell's stress tensor can subsequently be extracted from the lesser Green's function, which governs the field correlations. 
The final radiation formulas can be expressed in terms of Fresnel coefficients, revealing an insightful connection to energy conservation via Kirchhoff's law. 
Using the Wigner transform, the analytical expression for the radiative torque can also be related to the generalized Fresnel coefficients. 
Numerical calculations based on the optical conductivity of bismuth are presented to corroborate the analytical results. 
These results provide a unified framework for energy, momentum, and angular momentum radiation in gyrotropic thin films.
\end{abstract}

\maketitle

\section{Introduction}

Radiation is a fundamental mechanism by which energy and information are exchanged between matter and the electromagnetic field. 
Foundational contributions by Kirchhoff~\cite{Greffet_2024_Kirchhoff}, Stefan, and Boltzmann established key principles of thermal radiation, which were later unified with energy quantization in Planck’s law~\cite{Planck_1901}.
At present, heat transport continues to underpin technologies ranging from thermophotovoltaic systems~\cite{zhao_high-performance_2017,lapotin_thermophotovoltaic_2022} to nanoscale thermal management~\cite{latella_smart_2021, cuevas_radiative_2018}, motivating new formulations that extend beyond scalar energy flux.
In particular, the angular momentum carried by photons has emerged as a key degree of freedom, with relevance to quantum communication~\cite{wang_terabit_2012}, optical manipulation~\cite{parkin_measurement_2006, genet_chiral_2022}, and topological insulators~\cite{nogueira_fractional_2018}. 
These developments motivate a unified and analytically tractable description of radiative transport that incorporates energy and angular momentum fluxes. 

Analytical formulations of radiative transport have evolved significantly across classical and quantum domains. 
Planck’s radiation law introduced quantized energy exchange, enabling spectral resolution of energy flux. 
Subsequent developments incorporated momentum and angular momentum transport, with electromagnetic (EM) field formulations~\cite{Jackson_1999,Griffiths} and more recent extensions using Green’s functions (GFs) and fluctuational electrodynamics~\cite{polder_theory_1971, volokitin_radiative_2001, biehs_thermal_2007, narayanaswamy_greens_2014}. 
In low-dimensional systems and nanostructured media, radiation phenomena have been studied using surface mode dispersion relations~\cite{joulain_surface_2005}, and non-equilibrium approaches via the Keldysh formalism~\cite{richter_generalized_2008, wjs_transport}. 

In contrast to the extensively studied energy transport, angular momentum radiation remains comparatively less explored. 
Radiated electromagnetic fields can carry angular momentum, which has been experimentally demonstrated and measured~\cite{beth_mechanical_1936}, including radiation from chiral nanostructures~\cite{konishi_circularly_2011,nguyen_large_2023}. 
From a theoretical perspective, magneto‑optic effects~\cite{kollyukh_circular_2007,Sato_drude-model_2022} lift the degeneracy between left- and right-handed modes, enabling net angular momentum radiation. 
This is the approach adopted in the present study. 
Alternatively, the Haldane model can be used to generate angular momentum radiation without relying on an external magnetic field, leading to an analytical formula for angular momentum radiation~\cite{zym}.
However, the Born approximation was used in the previous work, where the photon GF was approximated by the vacuum version, thus not fully accounting for the presence of the material. 

In this work, we derive analytical expressions for the radiative transport of energy, linear momentum, and angular momentum while fully accounting for the interaction between the electromagnetic field and matter via the Dyson equation. 
For the case of a single metal film, the photon GFs can be solved exactly if the film is infinitesimally thin and the electron fluctuations in the out-of-plane direction can be neglected. 
Expressing the photon GFs in terms of Fresnel coefficients provides a natural interpretation in terms of the scattering of s- and p-polarized waves.
The EM field correlators can be obtained from the lesser photon GF in the Keldysh formalism, which in turn allows us to compute the various radiation quantities. 
The rate of energy emission can be recast in terms of Fresnel coefficients, elucidating its consistency with energy conservation as embodied in Kirchhoff's law.
The paper is structured as follows:
Section~\ref{sec:define} briefly introduces the Keldysh formalism and defines the photon GFs.
Section~\ref{sec:material} derives the retarded self-energy in the Dyson equation from the material’s frequency-dependent permittivity, an experimentally accessible quantity.
Section~\ref{sec:solve} presents an exact solution for the Dyson equation and analytical expressions for the resulting GFs.
Section~\ref{sec:fresnel} discusses the Fresnel coefficients of the system and how these can be used to represent the GFs.
Section~\ref{sec:transport} demonstrates how these GFs yield analytical expressions for radiative power, force, and torque.
Section~\ref{sec:numerical} applies the formalism to the Drude metal bismuth (Bi), with numerical plots illustrating the emission spectra.
Finally, Section~\ref{sec:conclusion} summarizes the discussion and outlines concluding remarks.

\section{Photon Green's Function} \label{sec:define}

We work exclusively in the temporal gauge~\cite{kay_quantum_2021}, where the electric field is given by $\vec{E} = - \partial_t \vec{A}$ and the magnetic field is given by $\vec{B} = \bm\nabla \times \vec{A}$. 
The electromagnetic fields are treated quantum mechanically in the Heisenberg picture, where the vector potential is an operator with time dependence.
The photon GF is then defined on the Keldysh contour~\cite{keldysh_diagram_2023, stefanucci_nonequilibrium_2013} as
\begin{equation}
    D_{\mu\nu}(\vec{r}, \tau; \vec{r}', \tau') = \frac{1}{i \hbar} \langle
    \mathcal{T}_C A_{\mu}(\vec{r}, \tau) A_{\nu}(\vec{r}', \tau')
    \rangle,
\end{equation}
where $\langle \ldots \rangle$ refers to the quantum expectation value~\cite{Dyson,Kadanoff_2018_QSM}. 
Contour times are represented by Greek symbols, and $\mathcal{T}_C$ denotes contour ordering. 
The indices $\mu$ and $\nu$ indicate the directions ($x$, $y$, $z$), so that $D_{\mu\nu}$ can be interpreted as components of a $3 \times 3$ matrix. 
$\tau$ and $\tau'$ are time variables on the Keldysh contour, and $\mathcal{T}_C$ is the contour-ordering superoperator ~\cite{wjs_transport,Rahman-2018-NEGF,stefanucci_nonequilibrium_2013}.
To obtain the lesser and greater versions of the GFs, which encode the distribution and correlation properties of the fields, we need to choose the contour times to lie on different branches, 
\begin{subequations}
\begin{align}
\begin{split}
    D_{\mu\nu}^<(\vec{r}, t, \vec{r}', t') &= \frac{1}{i \hbar} \langle
    A_{\nu}(\vec{r}', t') A_{\mu}(\vec{r}, t) 
    \rangle ,
\end{split} \label{eq:D_less} \\
\begin{split}
    D_{\mu\nu}^>(\vec{r}, t, \vec{r}', t') &= \frac{1}{i \hbar} \langle
    A_{\mu}(\vec{r}, t) A_{\nu}(\vec{r}', t') 
    \rangle \text{.}
\end{split} \label{eq:D_greater}
\end{align}
\end{subequations}

The retarded GF, which describes the causal response, is given by $D^R = \theta(t-t') (D^> - D^<)$~\cite{henneberger_exact_2009}.
It is obtained by solving the Dyson equation~\cite{abrikosov_methods_2012}, $D^R = v^R + v^R \Pi^R D^R$, or in full, 
\begin{align}
    &D^R(\vec{r}, t, \vec{r}', t') = v^R(\vec{r}-\vec{r}', t-t') \nonumber\\
    &+\int \d \vec{r}_1 \int \d t_1 \int \d \vec{r}_2\int \d t_2 \,v^R(\vec{r}-\vec{r}_1, t-t_1) \nonumber\\
    &\times \Pi^R(\vec{r}_1, t_1, \vec{r}_2, t_2) D^R(\vec{r}_2, t_2, \vec{r}', t') .
\end{align} \label{eq:Dyson_equation}
Here, $\Pi^R$ is the self-energy which encapsulates the material information, and $v^R$ is the vacuum version of $D^R$, which satisfies the GF equation $v^{-1} v^R(\vec{r}, t) = \delta^{(3)}(\vec{r}) \delta(t) I$.
The differential operator $v^{-1}$ is defined by $v^{-1} \vec{A} \equiv - \mu_0^{-1} (\nabla \times \nabla \times \vec{A} + c^{-2} \partial_t^2 \vec{A}) $, and its definition originates from one of the Maxwell's equations, $v^{-1} \vec{A} = - \vec{J}$.

Given the planar geometry of our system, we employ the following Fourier transform convention, 
\begin{align}
        D(z, z', \omega, \vec{k}_\parallel) =& \int \d t\, e^{i \omega (t-t') } \int \d^2 \vec{r}_\parallel \nonumber\\
        &\times e^{-i \vec{k}_\parallel \cdot (\vec{r}_\parallel-\vec{r}'_\parallel)}  D(\vec{r}, t, \vec{r}', t'),\label{eq:Fourier_definition}
\end{align}
where $\omega$ is the (angular) frequency and $\vec{k}_\parallel = k_x \hat{\vec{x}} + k_y \hat{\vec{y}}$ is the in-plane wavevector, with its magnitude denoted by $k_\parallel$. 
Similarly, $\vec{r}_\parallel = x \hat{\vec{x}} + y \hat{\vec{y}}$ is the in-plane position vector. 
The retarded solution for the vacuum GF in the mixed representation is~\cite{Keller}
\begin{widetext}
\begin{equation}
    v^R(z,\omega,k_x,k_y) 
    = \frac{\mu_0}{2i\gamma}  e^{i \gamma |z|} \biggl[
    \begin{pmatrix}
        1 & 0 & 0 \\
        0 & 1 & 0 \\
        0 & 0 & 0
    \end{pmatrix}
    - \frac{c^2}{\omega^2}
    \begin{pmatrix}
        k_x^2   & k_x k_y & k_x \gamma \sgn(z) \\
        k_x k_y & k_y^2   & k_y \gamma \sgn(z) \\
        k_x \gamma \sgn(z) & k_y \gamma \sgn(z) & -k_\parallel^2 - 2i\gamma\delta(z)
    \end{pmatrix}
    \biggr] \text{,}\label{eq:vr}
\end{equation}
\end{widetext}
where $\sgn$ is the signum function.
The longitudinal wavevector is given by $\gamma = (\omega^2/c^2 - k_\parallel^2)^{\frac{1}{2}}$ for the propagating case and $\gamma = i (k_\parallel^2 - \omega^2/c^2)^{\frac{1}{2}}$ for the evanescent case.

\section{Material-specific self-energy} \label{sec:material}

We provide a consistent expression for the self-energy function in terms of the material properties. 
From linear response theory~\cite{Kubo-1966,Golyk_linear-response_2013}, the photon self-energy in the Dyson equation can be interpreted as a current-current correlation,
\begin{equation}
    \Pi_{\mu\nu}(\bm{r},\tau;\bm{r}',\tau')
    = \frac{1}{i\hbar}\bigl\langle \mathcal{T}_C \xi_{\mu}(\bm{r},\tau) \xi_{\nu}(\bm{r}',\tau')\bigr\rangle .
\end{equation}
We use the symbol $\vec{\xi}$ for the random current, in contrast to the total current previously denoted by $\vec{J}$. 
The system under consideration is a two-dimensional thin film with a sufficiently large cross-sectional area ($\Sigma$) in the $x$ and $y$ directions, but with a small thickness ($L_z$) in the $z$ direction.
Because the system is confined to a plane, defined here as $z=0$, the current fluctuations possess only in‑plane components.
The retarded self-energy after the Fourier transform thus takes the form of 
\begin{subequations}
\begin{align}
    \Pi^R(z, z', \omega, \vec{k}_\parallel) &= \delta(z) \delta(z') \Pi_0(\omega, \vec{k}_\parallel), \label{eq:pi_delta_z} \\
    \Pi_0(\omega, \vec{k}_\parallel) &= \begin{pmatrix}
        \pi_{xx} & \pi_{xy} & 0 \\
        \pi_{yx} & \pi_{yy} & 0 \\
        0 & 0 & 0
    \end{pmatrix} . \label{eq:pi_no_z}
\end{align}
\end{subequations}
The differential form of the Dyson equation, written as $v^{-1} D^R = I + \Pi^R D^R$, can be recast into the more familiar form of $- \mu_0^{-1} \nabla \times \nabla \times D^R + \epsilon_0 \epsilon \omega^2 D^R = I$~\cite{van_vlack_spontaneous_2012, benisty_nanophotonics_2022}.
In the latter formulation, the material properties are described via the permittivity ($\epsilon$) rather than the self-energy.
Their equivalence emerges upon considering the induced current as governed by Ohm’s law.
Furthermore, the electric susceptibility ($\chi_e = \epsilon - 1$) is related to the 2D conductivity ($\sigma$) by $\sigma/L_z = -i \omega \epsilon_0 \chi_\textrm{e}$.

In the homogeneous and isotropic case, we have the simple relations $\pi_{xx} = \pi_{yy} = - L_z \epsilon_0 \chi_\textrm{e} \omega^2$~\cite{abrikosov_methods_2012} and $\pi_{xy} = \pi_{yx} = 0$. 
To enable radiation carrying angular momentum, a magnetic field ($\vec{B} = B \hat{\vec{z}}$) is applied to break reciprocity, thereby introducing gyrotropic terms into the self-energy, 
\begin{equation}
    \Pi_0(\omega,\bm{k}_\parallel)
    = \pi_0
    \begin{pmatrix}
        1 & \alpha & 0 \\
        -\alpha & 1 & 0 \\
        0 & 0 & 0
    \end{pmatrix} . \label{eq:pi_magnetic}
\end{equation}
The ratio of off-diagonal to diagonal terms ($\alpha$) scales linearly with $B$.
To obtain an exact expression for $\alpha$, we adopt the Drude model as a minimal yet effective description of metallic systems. 
Drude conductivity is derived by modeling charge carriers as classical particles that accelerate under an electric field and scatter randomly with a characteristic relaxation time. 
In the presence of a magnetic field, the conductivity tensor elements are slightly modified as~\cite{Sato_drude-model_2022}
\begin{subequations}
    \begin{align}
    \begin{split}
        \sigma_{xx}=\sigma_{yy} &= L_z \epsilon_0 \omega_p^2 \frac{(1-i\omega\tau) \tau}{(1-i\omega\tau)^2+(\omega_c\tau)^2},
    \end{split}\\
    \begin{split} 
        \sigma_{xy}=-\sigma_{yx} &= L_z \epsilon_0 \omega_p^2 \frac{\omega_c \tau^2}{(1-i\omega\tau)^2+(\omega_c\tau)^2},
    \end{split} 
     \end{align}
\end{subequations}
where $\tau$ is the Drude relaxation time, $\omega_c = e B / m^*_e$ is the cyclotron frequency, and $\omega_p=(n_ve^2/\epsilon_0 m^*_{e})^{\frac{1}{2}}$ is the plasma frequency~\cite{Kittel-2018-solidstate}. 
The other constants are: $n_v$ (electron volume density), $e$ (elementary charge), and $m_e^*$ (effective electron mass at conduction band).
Thus, the entries in the self-energy are given by
\begin{subequations}
\begin{align}
    \alpha &= \frac{\omega_c\tau}{1-i\omega\tau} , \label{eq:alpha} \\
    \pi_0 &= -L_z \epsilon_0\omega_p^2\frac{i\omega\tau}{1-i\omega\tau}\frac{1}{1+\alpha^2} . \label{eq:pi_0}
\end{align}
\end{subequations}
We identify bismuth (Bi) as a suitable candidate because of the availability of computational data and to its established role in magneto‑optical phenomena~\cite{Timrov_2012_bismuth,Scott_Magnetooptic_1976}. 
Although our system is a thin film, we employ parameter values corresponding to a bulk sample for calculation. 
This simplification is motivated by the fact that bulk data are more readily available and provide a reliable baseline, while thin‑film corrections can be incorporated in future work once the proof‑of‑concept is established.
In Table~\ref{tab:frequencies}, the three relevant frequency parameters for Bi are shown, all of which fall within the infrared region of the EM spectrum. 
\begin{table}[tb]
\caption{Parameters for Drude model of Bi~\cite{Kittel-2018-solidstate,Timrov_2012_bismuth}. \label{tab:frequencies}}
\begin{ruledtabular}
\begin{tabular}{cc}
Plasma frequency $\omega_p$ & \SI{8.51e14}{\per\second} \\
Drude damping $1/\tau$ & \SI{5.62e13}{\per\second} \\
Cyclotron frequency (\SI{10}{\tesla}) $\omega_c$ & \SI{2.93e14}{\per\second} \\
Effective electron mass $m^*_e$ & $0.006\,m_e$ \\
\end{tabular}
\end{ruledtabular}
\end{table}

\section{Solving the Green's functions} \label{sec:solve}

For isotropic and homogeneous materials, transport quantities such as energy, linear momentum, and angular momentum remain invariant under rotation about the $z$ axis. 
It is often more convenient to work in a rotated coordinate basis defined by the direction of $\vec{k}_\parallel$.
Following Sipe~\cite{Sipe}, we use the coordinate system defined by the right-handed triplet $(\hat{\vec{k}}_\parallel,\hat{\vec{s}}, \hat{\vec{z}})$, where the unit vectors are constructed by $\hat{\vec{k}}_\parallel = \vec{k}_\parallel/k_\parallel$ and $\hat{\vec{s}} = \hat{\vec{z}} \times \hat{\vec{k}}_\parallel$. 
The various matrix quantities typically take on simpler forms in this rotated basis, and henceforth, all matrix expressions will be presented in this new ordered basis unless otherwise specified. 
In other words, we are free to redefine the in-plane directions as $\hat{\vec{x}} = \hat{\vec{k}}_\parallel$ and $\hat{\vec{y}} = \hat{\vec{s}}$. 
Note that the form of $\Pi_0$ given in Eq.~\eqref{eq:pi_magnetic} is invariant under such a rotation.
For other quantities such as $v^R$, a convenient shortcut to obtain the rotated form of Eq.~\eqref{eq:vr} is to set $k_x=k_\parallel$ and $k_y=0$, rather than explicitly multiplying the rotation matrices.
In wave optics, the directions defined by s polarization (transverse electric, TE) and p polarization (transverse magnetic, TM) form another commonly used orthonormal basis, $(\hat{\vec{p}},\hat{\vec{s}}, \hat{\vec{k}})$.
The three-dimensional wavevector is defined as $\vec{k} = \vec{k}_\parallel + \gamma\sgn(z) \hat{\vec{z}}$. 
With the definition of the wavevector established, the p polarization direction is then defined by the orthonormal condition, $\hat{\vec{p}} = \hat{\bm{s}}\times\hat{\bm{k}}$.
As we are interested in the far-field regime, the evanescent modes decay exponentially, and only the propagating modes persist. 
Mathematically, this behavior is reflected in the GF by the exponential factor $e^{i \gamma z}$. 
For a mode to be propagating, it suffices to consider only the case where $\gamma$ is real, i.e., $\gamma^* = \gamma$. 

The Dyson equation, Eq.~\eqref{eq:Dyson_equation}, formally admits an infinite series solution,
\begin{align}
    D^R &= v^R + v^R \Pi^R v^R + v^R \Pi^R v^R \Pi^R v^R + \ldots \notag \\
    &= v^R (I - \Pi^R v^R)^{-1} . \label{eq:Dyson_series}
\end{align}
In the mixed representation, because of the presence of the Dirac delta functions in $\Pi^R$, one only needs to evaluate $v^R$ at $z=0$ for the $z$ convolutions in Eq.~\eqref{eq:Dyson_series}. 
We define $v_0$ as $v^R$ evaluated at $z-z'=0$, and without any $z$ components,
\begin{equation}
    v_0 = \frac{\mu_0}{2i\gamma} \biggl( \frac{c^2}{\omega^2} \gamma^2 \hat{\vec{k}}_\parallel \hat{\vec{k}}_\parallel + \hat{\vec{s}}\hat{\vec{s}} \biggr) 
    . \label{eq:v0}
\end{equation}
Another complication arises from the delta function in $v^R$ itself, which motivates the use of the thin-film approximation, such that the delta function is ignored when multiplied by $\Pi^R$ of the form in Eq.~\eqref{eq:pi_no_z}. 
Therefore, Eq.~\eqref{eq:Dyson_series} for the thin-film scenario simplifies to
\begin{equation}
    D^R(z,z') = v^R(z-z') + v^R(z) \Pi_0\mathcal{T} v^R(-z') , \label{eq:Dyson_solution}
\end{equation}
where $\mathcal{T}= (I - v_0 \Pi_0)^{-1}$ carries the interpretation of Fresnel transmission coefficients.
With the retarded GF worked out, the advanced GF is effortlessly obtained by $D^A_{\mu\nu}(\vec{r}, \vec{r}', \omega) = D^R_{\nu\mu}(\vec{r}', \vec{r}, \omega)^*$. 

The lesser GF, which encodes the field correlations as shown in Eq.~\eqref{eq:D_less}, can then be obtained by the Keldysh equation~\cite{kruger_trace_2012, aeberhard_photon_2014, wjs_transport} as
\begin{equation}
    D^<(z,z') = \int \d z_1 \int \d z_2 \, D^R(z, z_1) \Pi^<(z_1, z_2) D^A(z_2, z') . \label{eq:keldysh_equation}
\end{equation}
Here, the material is assumed to be at thermal equilibrium with a reservoir at temperature $T$.
Thus, the fluctuation-dissipation theorem applies, which states that the lesser self-energy is given by $\Pi^< = N(\omega) (\Pi^R - \Pi^A)$, where $N(\omega) = [\exp\bigl(\hbar\omega/(k_B T)\bigr) - 1]^{-1}$ is the Bose-Einstein function. 
Similar to the case for photon GF, the advanced form of the self-energy is $\Pi^A(z,z') = \Pi^R(z',z)^\dagger$. 
Although the full solution of retarded GF is available, only formula $D^R(z,0) = v^R(z) \widetilde{\mathcal{T}}$ is needed for subsequent calculations, where $\widetilde{\mathcal{T}}$ is also associated with the Fresnel coefficients and is defined as $\widetilde{\mathcal{T}}=(I-\Pi_0v_0)^{-1}$.
Therefore, the lesser GF, Eq.~\eqref{eq:keldysh_equation}, can be simplified as
\begin{equation}
    D^<(z,z') = N(\omega) v^R(z) \widetilde{\mathcal{T}} (\Pi_0 - \Pi_0^\dagger) \widetilde{\mathcal{T}}^\dagger v^R(z')^\dagger . \label{eq:Dless_thin}
\end{equation} 
Quantities involving two field operators, such as the Poynting vector or the Maxwell stress tensor, emerge naturally from $D^<$. 
For the $B=0$ case, $\Pi_0 = \pi_0$ can be treated as a scalar quantity. 
We then have $\Pi_0 - \Pi_0^\dagger = -2i L_z \Im(\epsilon_0 \epsilon) \omega^2$, and the role of $\Im(\epsilon)$ in thermal radiation is well-known in the fluctuational electrodynamics literature~\cite{francoeur_role_2008,joulain_surface_2005}.

\section{Fresnel Coefficients}\label{sec:fresnel}

Motivated by Kirchhoff's radiation law, we seek to formulate the lesser GF in terms of the Fresnel coefficients to connect material response with scattering theory~\cite{Sipe}.
Here, we justify why $\mathcal{T}$ and $\widetilde{\mathcal{T}}$ can be interpreted as the transmission coefficients.
For the simplest case of a conducting sheet with 2D conductivity $\sigma$ (without magnetic field), the Fresnel transmission coefficients are~\cite{inampudi_fresnel_2016}
\begin{subequations}
\begin{align}
    t_s =& [1 + \mu_0 \omega \sigma/(2\gamma)]^{-1} , \label{eqn:ts} \\
    t_p =& [1 + \gamma \sigma/(2 \epsilon_0 \omega)]^{-1} . \label{eqn:tp}
\end{align}
\end{subequations}
Our sign convention for the transmission coefficients allows us to express the corresponding reflection coefficients as simply $r_{s/p} = t_{s/p} - 1$. 
Thus, it can be verified that $\mathcal{T}=\widetilde{\mathcal{T}} =  t_p \hat{\vec{k}}_\parallel \hat{\vec{k}}_\parallel + t_s \hat{\vec{s}} \hat{\vec{s}} + \hat{\vec{z}} \hat{\vec{z}}$ is in accordance with its interpretation as the transmission matrix. 
In the presence of a magnetic field, $\mathcal{T}$ and $\widetilde{\mathcal{T}}$ acquire off-diagonal terms. 
In the $(\hat{\vec{k}}_\parallel,\hat{\vec{s}}, \hat{\vec{z}})$ basis, they are related by
\begin{subequations}
    \begin{align}
    &\mathcal{T} =(I-v_0\Pi_0)^{-1}
    = \begin{pmatrix}
        t_{pp} & t_{ps} & 0 \\
        t_{sp} & t_{ss} & 0 \\
        0 & 0 & 1
    \end{pmatrix} ,\label{eq:define_t} \\
    &\widetilde{\mathcal{T}} =(I-\Pi_0 v_0)^{-1}
    = \begin{pmatrix}
        t_{pp} & -t_{sp} & 0 \\
        -t_{ps} & t_{ss} & 0 \\
        0 & 0 & 1
    \end{pmatrix} . 
    \end{align}
\end{subequations}
The matrix elements $t_{mn}$ represent the transmission probability for a photon initially in polarization state $n$ to emerge in polarization state $m$. 
In the usual case with $B=0$, s polarized and p polarized light propagate independently and do not mix. 
In the presence of a magnetic field, the permittivity tensor becomes gyrotropic with the emergence of cross-polarization terms, $t_{sp}$ and $t_{ps}$, enabling polarization rotation via the polar Kerr effect~\cite{Scott_Magnetooptic_1976}. 
By evaluating Eq.~\eqref{eq:define_t}, the relation $t_{ps} = -(c \gamma/\omega)^2 t_{sp}$ emerges, allowing one of them to be eliminated.
Eq.~\eqref{eq:define_t} can also be written recursively as $\mathcal{T} - I = v_0 \Pi_0 \mathcal{T}$, establishing important relations between the remaining elements without having to solve for them explicitly, 
\begin{align}
    \begin{pmatrix}
        t_{pp}-1 & -\frac{c^2 \gamma^2}{\omega^2} t_{sp} \\
        t_{sp} & t_{ss} - 1
    \end{pmatrix}
    = &\frac{\mu_0 \pi_0}{2 i \gamma}
    \begin{pmatrix}
        \frac{c^2 \gamma^2}{\omega^2} & \alpha \frac{c^2 \gamma^2}{\omega^2} \\
        -\alpha  & 1
    \end{pmatrix} \nonumber\\
    &\times
    \begin{pmatrix}
        t_{pp} & -\frac{c^2 \gamma^2}{\omega^2}t_{sp} \\
        t_{sp} & t_{ss}
    \end{pmatrix} .
\end{align}
The generalized Fresnel coefficients $t_{ss}$, $t_{pp}$, and $t_{sp}$ remain expressed in terms of the self-energy parameters $\alpha$ and $\pi_0$. 

Next, we substitute the photon vacuum GF $v^R$ and $\widetilde{\mathcal{T}}$ under the new basis into Eq.~\eqref{eq:Dless_thin}, which leads to
\begin{align}
    D^<(z, z') 
    =& \frac{\mu^2_{0}}{2\gamma^2} N(\omega) e^{i\gamma(|z|-|z'|)} \nonumber\\
    &\times \biggl[  i \Im(\pi_0) F_1 + \Re(\alpha\pi_0) F_2  \biggr] , \label{fo:D-lesser}
\end{align}
with $F_1$ and $F_2$ given by
\begin{subequations}
\begin{align}
    F_1 \sim
    & \begin{pmatrix}
       \frac{c^4\gamma^4}{\omega^4}|t_{sp}|^2+\frac{c^4\gamma^4}{\omega^4}|t_{pp}|^2  & -\frac{c^2\gamma^2}{\omega^2}t_{ss}^*t_{sp}+\frac{c^4\gamma^4}{\omega^4}t_{pp}t^*_{sp} \nonumber\\
       -\frac{c^2\gamma^2}{\omega^2}t_{ss}t_{sp}^*+\frac{c^4\gamma^4}{\omega^4}t_{pp}^*t_{sp}  & |t_{ss}|^2+\frac{c^4\gamma^4}{\omega^4}|t_{sp}|^2
    \end{pmatrix} , \\ 
    F_2 \sim
    & \begin{pmatrix}
        -\frac{c^4\gamma^4}{\omega^4}t_{pp}t^*_{sp} + \frac{c^4\gamma^4}{\omega^4}t^*_{pp}t_{sp} & \frac{c^2\gamma^2}{\omega^2}t_{ss}^*t_{pp} + \frac{c^4\gamma^4}{\omega^4}|t_{sp}|^2\\
        -\frac{c^2\gamma^2}{\omega^2}t_{ss}t^*_{pp}-\frac{c^4\gamma^4}{\omega^4}|t_{sp}|^2 & -\frac{c^2\gamma^2}{\omega^2}t_{ss}t^*_{sp}+\frac{c^2\gamma^2}{\omega^2}t^*_{ss}t_{sp}
    \end{pmatrix} .
\end{align}
\end{subequations}
Although $F_1$ and $F_2$ are technically $3 \times 3$ quantities, we only spell out the $x$ and $y$ components for brevity. 
There is a simple relationship to determine any $z$ components by
\begin{equation}
    D^<_{z\mu}=-\frac{k_\parallel}{\gamma}\sgn(z)D^<_{x\mu} , \label{eq:z_component_dless}
\end{equation}
for $\mu =x, y, z$.

\section{Radiative transport} \label{sec:transport}

In classical electromagnetism, the energy flux density is given by the Poynting vector, $\vec{S} = \mu_0^{-1} \vec{E} \times \vec{B}$~\cite{Griffiths,Jackson_1999}. 
In the present geometry, only the $z$-component is nonzero because of symmetry. 
In the quantum formulation, careful attention must be paid to operator ordering, as well as to contributions arising from zero-point motion. 
The standard approach in quantum field theory involves employing operator normal ordering to systematically eliminate vacuum divergences~\cite{Agarwal_black-body-fluctuations_1975}. 
Thus, we use $\langle S_z \rangle = \mu_0^{-1} \langle : E_x B_y - E_y B_x : \rangle$.
The lesser GF is the appropriate choice as it yields the $N(\omega)$ prefactor essential for reproducing Planck's radiation formula. 
The effect of normal ordering can be achieved by integrating only the positive-frequency part of the spectrum, multiplying by 2, and taking the real part, which also has the desired effect of symmetrizing the expressions~\cite{wjs_transport,Yap-2017}. 
For example, we can calculate the normal-ordered correlation $\langle : E_x B_y : \rangle$ by
\begin{align}
    &\Big\langle :E_x(\vec{r},t) B_y(\vec{r}',t'): \Big\rangle \nonumber\\
    =& 2\Re \int_{0}^\infty \frac{\d \omega}{2\pi} 
    \int \frac{\d^2 \vec{k}_\parallel}{(2\pi)^2} \, e^{-i \omega (t-t') +i \bm{k}_\parallel \cdot (\vec{r}_\parallel -\vec{r'}_\parallel)} \nonumber\\
    &\times \hbar \omega [ \partial_z D^<_{xx}(\omega, \vec{k}_\parallel) - i k_x D^<_{zx}(\omega, \vec{k}_\parallel) ] .
\end{align}
From the form of Eq.~\eqref{fo:D-lesser}, the $z$ derivatives are evaluated using the rules $\partial_z\to i\gamma\sgn(z)$ and $\partial_{z'}\to -i\gamma\sgn(z')$. 
We can also use Eq.~\eqref{eq:z_component_dless} to substitute $D^<_{zx}$ with $D^<_{xx}$. 
Although the lesser GF encodes non-local correlations in general, transport quantities only require its evaluation at coincident points.
Therefore, the exponential in the integrand vanishes when we require $\vec{r}' \to \vec{r}$ and $t' \to t$. 
It is helpful to keep the variables $\vec{r}$ and $\vec{r}'$ distinct until the end to avoid confusion when taking the partial derivatives. 
As noted previously, evanescent modes are exponentially suppressed in the far-field regime, and thus the $\vec{k}_\parallel$ integral is restricted to the propagating regions where $k_\parallel < \omega/c$. 
Upon simplifying, the expression for the Poynting vector is 
\begin{align}
    \langle S_z \rangle =& \frac{1}{\mu_0}\int_{0}^{\infty}\frac{\d\omega}{2\pi}~i\hbar\int_{k_\parallel < 
    \frac{\omega}{c}}\frac{\d^{2}\bm{k}_\parallel}{(2\pi)^2} \notag \\
    &\times \frac{2\omega}{\gamma} \left(\gamma^2 D_{yy}^{<}+\frac{\omega^2}{c^2}D_{xx}^{<}\right) \notag \\ 
    =& \int_{0}^{\infty}\frac{\d\omega}{\pi}~\hbar \omega N(\omega) \int_{k_\parallel < 
    \frac{\omega}{c}}\frac{\d^{2}\bm{k}_\parallel}{(2\pi)^2} \notag \\
    &\times \Re \Tr [(I - \mathcal{T}) \widetilde{\mathcal{T}}^\dagger] \sgn(z). \label{eq:poynting_integrand}
\end{align}
The total power emitted is then proportional to the area of the film, denoted by $\langle I \rangle = \langle S_z \rangle \Sigma$. 
The signum function correctly indicates that the Poynting vector points away from the film surface even if we consider $z < 0$.
Since the focus is on emission to $z = + \infty$, $\sgn(z)=1$, and can be safely omitted from future expressions. 

To establish the connection with Kirchhoff's law, we define the integrand in Eq.~\eqref{eq:poynting_integrand} as 
\begin{equation}
    A_I = \Re \Tr [(I - \mathcal{T}) \widetilde{\mathcal{T}}^\dagger] .
\end{equation}
Expanded out in terms of transmission coefficients, this is 
\begin{align}
    A_{I} = \frac{1}{2} \bigl(2 - |t_{ss}-1|^2 - |t_{ss}|^2 - |t_{pp}-1|^2 - |t_{pp}|^2 \nonumber\\
    - 4 \frac{c^2\gamma^2}{\omega^2} |t_{sp}|^2 \bigr) . \label{eq:ai_fresnel}
\end{align}
In the special case where there is no applied magnetic field, the cross term vanishes, and $A_I$ simplifies to
\begin{equation}
    A_I = \frac{1}{2} \bigl( 2 - |r_s|^2 - |t_s|^2 - |r_p|^2 - |t_p|^2 \bigr) . 
\end{equation}
For a unit incident energy flux, $|r|^2$ and $|t|^2$ represent the fractions of power reflected and transmitted, respectively.
By energy conservation, the quantity $1-|r|^2-|t|^2$ represents the fraction of incident energy that is neither reflected nor transmitted, and therefore corresponds to absorption within the material. 
The s- and p-polarization can be regarded as independent degrees of freedom, each contributing equally, which accounts for the factor of one-half. 
We have thus established an indirect connection between the emission and absorption properties, embodying Kirchhoff's radiation law of detailed balance~\cite{Greffet_2024_Kirchhoff}. 
Similar findings are reported in Refs.~\cite{polder_theory_1971,joulain_surface_2005}, though their analysis pertains to a semi-infinite slab geometry, where there would be no transmission. 
For the case $B \neq 0$, Lorentz reciprocity is broken, and Kirchhoff’s law is no longer expected to hold.
Owing to the magneto-optic effect, a new term $|t_{sp}|^2$ emerges in Eq.~\eqref{eq:ai_fresnel}. 
However, certain generalizations of Kirchhoff's law have been recently proposed~\cite{guo_adjoint_2022}.

The treatment of linear momentum transport follows in close analogy to the analysis of energy emission. 
The radiative force, in the static case~\cite{Griffiths}, is given by $\langle N_z \rangle= -\langle T_{zz}\rangle \Sigma$, where $T_{zz}$ is the $zz$ component of Maxwell stress tensor, $\langle T_{zz} \rangle = \langle: \mu_0^{-1} (B_z^2 - B^2/2) + \epsilon_0 (E_z^2 - E^2/2):\rangle$. 
Expressed in terms of $D^<$, we have
\begin{align}
    \langle T_{zz}\rangle =& -\frac{1}{\mu_0}\int_{0}^{\infty}\frac{\d\omega}{2\pi}~i\hbar\int_{k_\parallel < \frac{\omega}{c}}\frac{\d^{2}\bm{k}_\parallel}{(2\pi)^2} \nonumber\\
    &\times 2 \left( \gamma^2D_{yy}^{<} + \frac{\omega^2}{c^2} D_{xx}^{<} \right) .
\end{align}
Compared to Eq.~\eqref{eq:poynting_integrand}, the only difference is a factor of $-\gamma/\omega$ in the integrand. 
This ties in with our understanding of the ratio of momentum to energy of an EM plane wave. 

For angular momentum transport, the radiation torque is also obtained from the Maxwell stress tensor by $\langle M_z \rangle = \langle:\textstyle\int (yT_{xz}-xT_{yz}) \d x \d y:\rangle$~\cite{zym,Stephen-2001-OAM}.
Since the integrand contains position coordinates, the integral cannot be directly evaluated into a flux quantity proportional to the area.
If the quantum expectation of the momentum component is evaluated before the coordinates integral, because of the $\bm{r}'_\parallel \to \bm{r}_\parallel$ step, the momentum component $\langle T_{xz} \rangle$ and $\langle T_{yz} \rangle$ will no longer contain $x$ and $y$ coordinates. 
In contrast, the result depends on geometry or potentially even be zero~\cite{Maghrebi-2019-torque,Stephen-2001-OAM}, in deviation from our expectations. 
To address this complication, we proceed using the Wigner transform method, which is detailed in Appendix~\ref{sec:wigner}. 
The final expression of radiative torque can still be expressed in the form of the product of flux density and area. 
Thus, we formally denote the angular momentum flux as $\langle yT_{xz}-xT_{yz}\rangle \Sigma$, where
\begin{align}
    \langle yT_{xz} -  xT_{yz}\rangle
    =& -\frac{1}{\mu_0}\int_{0}^{\infty}\frac{\d\omega}{2\pi}~\hbar\int_{k_\parallel < \frac{\omega}{c}}\frac{\d^{2}\bm{k}_\parallel}{(2\pi)^2} \notag \\
    &\times \gamma \left( D^{<}_{xy} -D^{<}_{yx} \right) \notag \\
    =& \int_{0}^{\infty}\frac{\d\omega}{\pi} N(\omega)\int_{k_\parallel < \frac{\omega}{c}}\frac{\d^{2}\bm{k}_\parallel}{(2\pi)^2} \hbar A_M .
\end{align}
Similar to the case of $A_I$, we seek to express $A_M$ in terms of the transmission coefficients, rather than elements of $D^<$. 
Upon simplification, $A_M$ is given by
\begin{equation}
    A_M = \frac{1}{2}\Im \Tr \biggr[(I-\mathcal{T} ) \widetilde{\mathcal{T}}^\dagger \biggr(\frac{c^2\gamma^2}{\omega^2}\hat{\bm{k}}_\parallel \hat{\bm{s}} -\hat{\bm{s}}\hat{\bm{k}}_\parallel\biggr)\biggr] .
\end{equation}
Alternatively, $A_M$ can be expanded in terms of the Fresnel coefficients as 
\begin{equation}
    A_{M} =
    \frac{c^2\gamma^2}{\omega^2}\Im [(t_{ss}+t_{pp}-1) t^*_{sp}] .
\end{equation}
Since $A_M$ is proportional to $t_{sp}^*$, the angular momentum radiated vanishes when there is no magnetic field.

We note that the diagonal terms of lesser GF, $D^{<}_{xx}$ and $D^{<}_{yy}$ are related to energy and linear momentum transport; on the other hand, the off-diagonal terms, $D^{<}_{xy}$ and $D^{<}_{yx}$ are related to angular momentum transport.
The analytical formulas for radiative power $\langle I\rangle$, force $\langle N_z\rangle$, and torque $\langle M_z\rangle$ can altogether be cast in a similar form, 
\begin{subequations}
\begin{align}
    \frac{\langle I \rangle}{\Sigma} &= \int^{\infty}_{0} \frac{\d\omega}{\pi}N(\omega)  \int_{k_\parallel < \frac{\omega}{c}} \frac{\d^{2}\bm{k}_\parallel}{(2\pi)^2} \hbar\omega A_{I}(k_\parallel,\omega) , \\
    \frac{\langle N_z \rangle}{\Sigma} &= \int^{\infty}_{0} \frac{\d\omega}{\pi}N(\omega) \int_{k_\parallel < \frac{\omega}{c}} \frac{\d^{2}\bm{k}_\parallel}{(2\pi)^2} \hbar\gamma  A_{I}(k_\parallel,\omega), \\
    \frac{\langle M_z \rangle}{\Sigma} &= \int^{\infty}_{0} \frac{\d\omega}{\pi}N(\omega) \int_{k_\parallel < \frac{\omega}{c}} \frac{\d^{2}\bm{k}_\parallel}{(2\pi)^2} \hbar A_{M}(k_\parallel,\omega) .
\end{align}
\end{subequations}
The physical interpretation of these formulas is straightforward, since $\hbar \omega$, $\hbar \gamma$, and $\hbar$ correspond to the energy, the $z$-component of linear momentum, and the angular momentum carried by a single photon.

\section{Numerical calculation} \label{sec:numerical}
\begin{figure}[!h]
    \centering
    \includegraphics[width=2.5in]{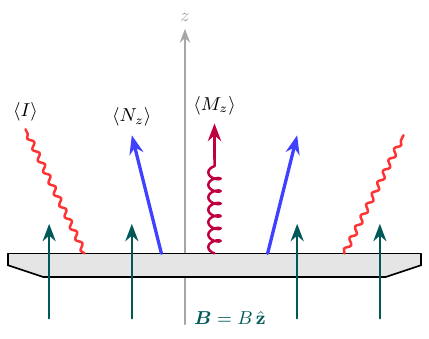}
    \caption{Schematic diagram of the system, where gray represents a 2D thin-film material applied with an out-of-plane magnetic field $\bm{B}$. The radiation quantities of interest are the power $(\langle I \rangle)$, force $(\langle N_z \rangle)$, and torque $(\langle M_z\rangle)$.}
    \label{fig.f1}
\end{figure}
%
\begin{figure}[tb]
    \centering
    \begin{tikzpicture}
        \node[anchor=south west,inner sep=0] (image) at (0,0)
            {\includegraphics[width=0.45\textwidth]{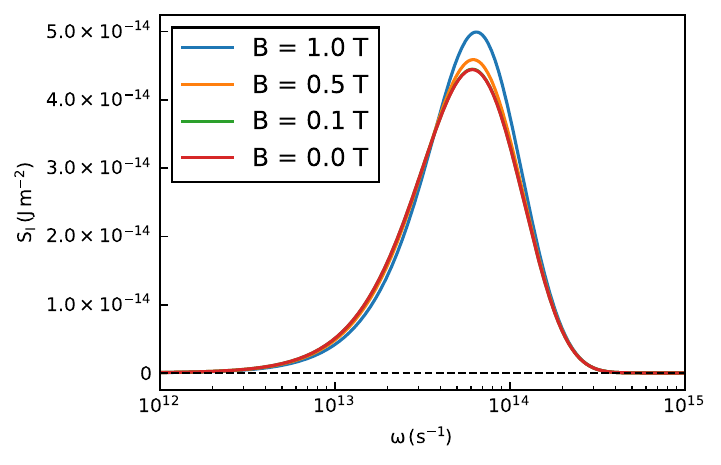}};
        \node[anchor=north west] at (image.north west) {(a)};
    \end{tikzpicture}
    \hfill
    \begin{tikzpicture}
        \node[anchor=south west,inner sep=0] (image) at (0,0)
            {\includegraphics[width=0.45\textwidth]{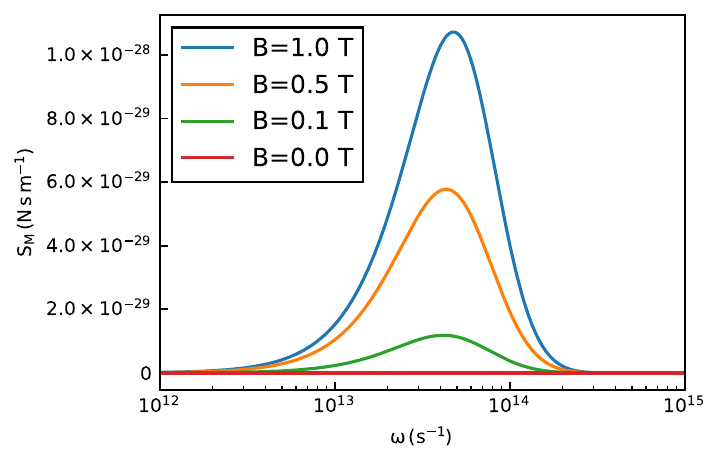}};
        \node[anchor=north west] at (image.north west) {(b)};
    \end{tikzpicture}

    \caption{Plots for (a) Power spectrum ($S_I$) and (b) torque spectrum ($S_M$) using the analytical formulas, Eqs. \eqref{eqn:define_si} and \eqref{eqn:define_sm}, respectively. Magnetic field used ranges from \SIrange{0}{1}{\tesla}. Drude parameters used are shown in Table \ref{tab:frequencies}. Room temperature (\SI{300}{\kelvin}) is used in all simulations. }
    \label{fig:spectrum_vary_b}
\end{figure}

Having established the analytical radiation formulas, it is instructive to examine a numerical example using a real material (Bi) as a reference. 
The Drude model parameters for Bi in Table~\ref{tab:frequencies} are used to calculate $\alpha$ and $\pi_0$ via Eq.~\eqref{eq:alpha} and Eq.~\eqref{eq:pi_0}, respectively.
They are subsequently used to evaluate the Fresnel coefficients defined in Eq.~\eqref{eq:define_t}.
The diagram is shown in Fig.~\ref{fig.f1}, where we set the thickness and the area of bismuth film as $L_z = \SI{1}{\nano\metre}$ and $\Sigma = \SI{1}{\centi\metre\squared}$.
Representative values of the total radiative power, force, and torque at various applied magnetic field strengths and at room temperature are presented in Table~\ref{tab:total_value}. 
For comparison, a perfect blackbody at \SI{300}{\kelvin} radiates a much larger power (\SI{459}{\watt\per\square\meter}) as given by the Stefan-Boltzmann law.

To facilitate a visualization of the frequency-resolved quantities, we identify the spectral functions as the integrands of the $\omega$ integrals,
\begin{subequations}
\begin{align}
    S_{I}(\omega) &= \frac{1}{\pi} N(\omega) \int_{k_\parallel < \frac{\omega}{c}} \frac{\d^2 \bm{k}_\parallel}{(2\pi)^2}\hbar\omega A_{I}(k_\parallel,\omega) , \label{eqn:define_si} \\
    S_{N}(\omega) &= \frac{1}{\pi} N(\omega) \int_{k_\parallel < \frac{\omega}{c}} \frac{\d^2 \bm{k}_\parallel}{(2\pi)^2} \hbar\gamma A_{I}(k_\parallel,\omega) , \\
    S_{M}(\omega) &= \frac{1}{\pi} N(\omega) \int_{k_\parallel < \frac{\omega}{c}} \frac{\d^2\bm{k}_\parallel}{(2\pi)^2}\hbar A_{M}(k_\parallel,\omega) . \label{eqn:define_sm}
\end{align}
\end{subequations}
\begin{table}[tb]
\caption{Radiative power, force, and torque at various magnetic field strengths and room temperature (\SI{300}{\kelvin}). The area of the film used is $\Sigma = \SI{1}{\centi\meter^2}$. } \label{tab:total_value}
\begin{ruledtabular}
\begin{tabular}{ccccc}
$B$ (\si{\tesla}) & 0 & 0.1 & 0.5 & 1.0 \\
\colrule
$\langle I \rangle$ ($10^{-6}$ \si{\watt}) & 511.12 & 511.57 & 522.22 & 553.11 \\
$\langle N_z \rangle$ ($10^{-15}$ \si{\newton}) & 990.03 & 990.91 & 1011.65 & 1071.82 \\
$\langle M_z \rangle$ ($10^{-21}$ \si{\newton\meter}) & 0 & $-$87.36 & $-$429.71 & $-$817.56 
\end{tabular}
\end{ruledtabular}
\end{table}
Given the common factors between the formulas for energy and linear momentum transport, we do not present separate plots for the linear momentum. 
Instead, we direct our attention to the angular momentum radiation, whose behavior is less intuitive and less commonly explored.
The results shown in Fig.~\ref{fig:spectrum_vary_b}(a) and \ref{fig:spectrum_vary_b}(b) correspond to the spectral functions of energy and angular momentum spectral density plotted against frequency.
The spectra, plotted on a logarithmic scale, are mostly concentrated between $\SI{e12}{\per\second}$ and $\SI{e15}{\per\second}$, corresponding to the infra-red range typical of objects at room temperature. 
For the power spectrum, the presence of a magnetic field has little influence on the overall shape of the curves. 
For the torque spectrum, the influence of the magnetic field is more pronounced, with the peak amplitudes scaling approximately in proportion to $B$. 
For both the power and torque spectra, the application of a weak magnetic field amplifies the peak magnitudes and shifts them toward higher frequencies. 
This behavior can be interpreted as the material exhibiting a higher effective temperature. 
Although the calculated values of torque are negative, we present the absolute values in Fig.~\ref{fig:spectrum_vary_b}(b). 
The negative sign simply indicates that the radiated torque has a direction opposite to the external magnetic field, consistent with previously reported results~\cite{zym, Stephen-2001-OAM}. 
Moreover, for emission to $z = - \infty$, the angular momentum radiated is also negative. 
In contrast to the radiated linear momentum (Poynting vector), which exhibits opposite signs on the two sides of the film, the angular momentum radiation retains the same sign on both sides. 
This behavior is expected, since the angular momentum is an odd function of $B$ and follows from the symmetry under $z \to -z$.
\begin{figure}[tb]
    \centering
    \includegraphics[width=\linewidth]{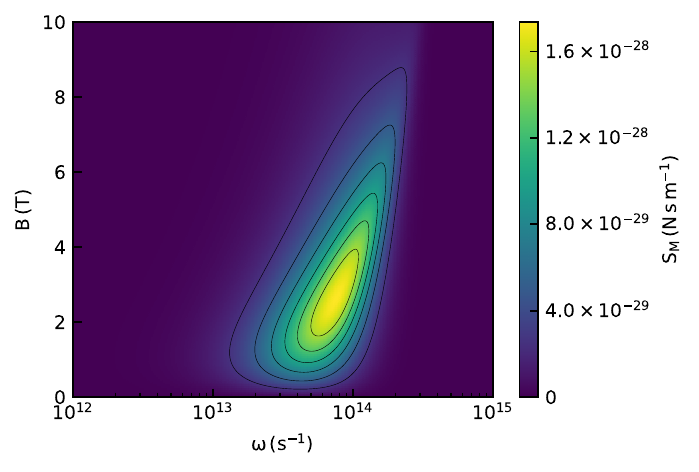}
    \caption{Heatmap to visualize how the angular momentum spectrum varies with the applied magnetic field. Each horizontal slice shows $S_M(\omega)$ [defined in Eq.~\eqref{eqn:define_sm}] at a particular level of $B$. }
    \label{fig:fM2}
\end{figure}
\begin{figure}[tb]
    \centering
    \includegraphics[width=\linewidth]{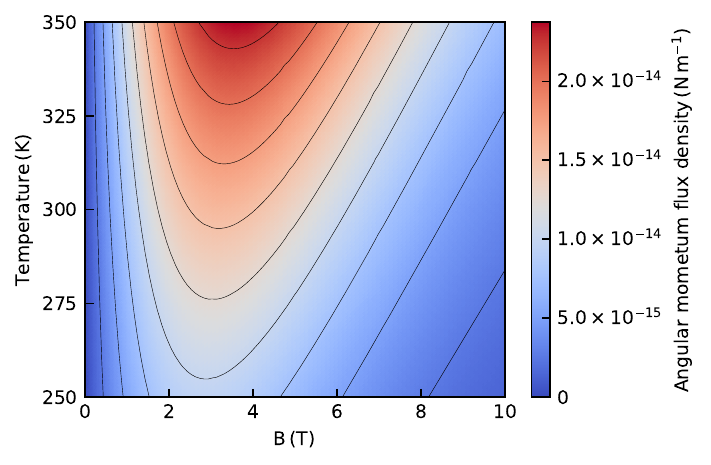}
    \caption{Heatmap to visualize the dependence of the total radiative torque ($\langle M_z \rangle$) on the applied magnetic field ($B$) and temperature ($T$). }
    \label{fig:fM3}
\end{figure}

Next, we examine the torque spectrum under a moderately stronger magnetic field.
Fig.~\ref{fig:fM2} shows the dependence of the torque spectrum $S_{M}$ on magnetic field strength up to \SI{10}{\tesla}. 
We observe that, at weak magnetic fields, the peak of $S_M$ approaches the Drude damping constant, whereas at strong fields it shifts toward the plasma frequency.
Fig.~\ref{fig:fM3} illustrates the variation of the angular momentum ($\langle M_z \rangle$), already integrated over frequency, with temperature and magnetic field. 
The temperature is scanned over a relatively mild range of $\SI{250}{\kelvin}$ to $\SI{350}{\kelvin}$, where no phase transition is expected. 
Intuitively, the intensity of the radiated power and torque is positively correlated with temperature. 
More interestingly, the total radiated torque does not increase monotonically with $B$, but instead exhibits a maximum around \qtyrange{3}{4}{\tesla}. 
The same conclusion can also be drawn from Fig.~\ref{fig:fM2} for a specific temperature. 

\section{Conclusion} \label{sec:conclusion}

To summarize, we have used the non-equilibrium GF method to obtain analytical formulas for the radiation of energy, momentum, and angular momentum. 
An analytical solution for Dyson's equation is possible by making the thin-film approximation.
Our radiation quantities are expressed in terms of generalized Fresnel coefficients, and Kirchhoff's law is recovered in the case of no magnetic field. 
More generally, to allow for angular momentum radiation, a magnetic field is introduced to break reciprocity, generating off-diagonal terms in the conductivity/permittivity tensor. 
To facilitate the derivation of angular momentum radiation, we applied the Wigner transform method to simplify integrals involving the GFs. 
Numerical calculations for the Drude metal Bi demonstrate that while the magnetic field modestly enhances radiated power, the associated torque notably exhibits a non-monotonic dependence on field strength, reaching a maximum near \qtyrange{3}{4}{\tesla} at ambient temperatures. 
These findings highlight the subtle interplay between magneto-optical effects and thermal radiation, suggesting opportunities for controlling angular momentum transfer in nanoscale photonic systems.

\section{Acknowledgments}

We acknowledge support from the Ministry of Education, Singapore, under the Academic Research Fund (FY2022).

\appendix

\section{Wigner Transform} \label{sec:wigner}

We introduce the Wigner transform~\cite{Wigner-1932, Kamenev-2011-FTofNEsystems,Golyk_linear-response_2013,Tim-2024-KBequations}, which is required to derive the expression for the radiative torque. 
For clarity, we consider only in-plane coordinates and omit any $z$ or $z'$ dependence here. 
The Fourier transform for translationally invariant functions is generally taken with respect to the difference of the two arguments,  
\begin{equation}
    D(\bm{k}_\parallel) = \int \d^2 \bm{r}_\parallel
    \, e^{-i \bm{k}_\parallel \cdot (\bm{r}_\parallel - \bm{r}'_\parallel)} D(\bm{r}_\parallel-\bm{r}_\parallel') \text{.} \label{eq:Fourier_definition_2}
\end{equation}
For notational purposes, we denote both a function and its Fourier transform with the same symbol, with the argument(s) indicating which version is intended.
In the absence of translational symmetry, for example, in a finite sample, we can perform independent Fourier transforms on each argument, adopting the convention that the exponents carry opposite signs,
\begin{align}
    D(\bm{k}_\parallel,\bm{k}'_\parallel) =& \int \d^2 \bm{r}_\parallel \int\d^2 \bm{r}'_\parallel e^{-i\bm{k}_\parallel \cdot \bm{r}_\parallel} \nonumber\\
    &\times e^{i \bm{k}_\parallel' \cdot \bm{r}_\parallel'} D(\bm{r}_\parallel,\bm{r}'_\parallel) \text{.}
\end{align}
In the Wigner transform, we use the average (A) and relative (R) coordinates. 
We define them as
\begin{subequations}
\begin{align}
    \bm{r}^A_\parallel &= (\bm{r}_\parallel + \bm{r}_\parallel')/2 \text{,} \\
    \bm{r}^R_\parallel &= \bm{r}_\parallel - \bm{r}'_\parallel \text{,} \\
    \bm{k}^A_\parallel &= (\bm{k}_\parallel+\bm{k}_\parallel')/2 \text{,} \\
    \bm{k}^R_\parallel &= \bm{k}_\parallel - \bm{k}_\parallel' \text{.} 
\end{align}
\end{subequations}
Thus, the double Fourier transform can also be expressed in terms of the relative-average coordinates as 
\begin{align}
    D(\bm{k}^R_\parallel, \bm{k}^A_\parallel)
    =& \int \d^2 \bm{r}_\parallel \int\d^2 \bm{r}'_\parallel
    \,e^{-i(\bm{k}^A_\parallel + \frac{1}{2} \bm{k}^R_\parallel) \cdot \bm{r}_\parallel} \nonumber\\
    &\times e^{i (\bm{k}^A_\parallel - \frac{1}{2} \bm{k}^R_\parallel) \cdot \bm{r}'_\parallel} D(\bm{r}_\parallel,\bm{r}_\parallel') \nonumber\\
    =& \int \d^2 \bm{r}^R_\parallel \int\d^2 \bm{r}^A_\parallel \nonumber\\
    &\times e^{-i \bm{k}^A_\parallel \cdot \bm{r}^R_\parallel} e^{-i \bm{k}^R_\parallel \cdot \bm{r}^A_\parallel} D(\bm{r}^R_\parallel,\bm{r}^A_\parallel) \text{.}
\end{align}
The Wigner transform is an incomplete Fourier transform of only one of the two arguments, where the resulting function uses the average coordinate $\bm{r}^A_\parallel$ and the average wavevector $\bm{k}^A_\parallel$, 
\begin{align}
    D(\bm{r}^A_\parallel, \bm{k}^A_\parallel) 
    &= \int \d^2 \bm{r}^R_\parallel e^{-i \bm{k}^A_\parallel \cdot \bm{r}^R_\parallel}  D(\bm{r}^R_\parallel,\bm{r}^A_\parallel) \text{.}
\end{align}
Similarly, we can also define the Wigner transform that uses the relative coordinates,
\begin{equation}
\begin{split}
    D(\bm{r}^R_\parallel, \bm{k}^R_\parallel)
    &= \int \d^2 \bm{r}^A_\parallel e^{-i \bm{k}^R_\parallel \cdot \bm{r}^A_\parallel}  D(\bm{r}^R_\parallel,\bm{r}^A_\parallel) \text{.}
\end{split}
\end{equation}
For translationally invariant functions, the GF will depend only on the relative position, i.e., $D(\bm{r}^R_\parallel,\bm{r}^A_\parallel)=D(\bm{r}^R_\parallel) =D(\bm{r}_\parallel-\bm{r}'_\parallel)$. In this representation, it is
\begin{subequations}
\begin{align}
    D(\bm{r}^A_\parallel,\bm{k}^A_\parallel) 
    &= \int \d^2 \bm{r}^R_\parallel \, e^{-i \bm{k}^A_\parallel \cdot \bm{r}^R_\parallel} D(\bm{r}^R_\parallel) \nonumber \\
    &= D(\bm{k}^A_\parallel) \text{,} \\
    D(\bm{r}^R_\parallel,\bm{k}^R_\parallel) 
    &= \int \d^2 \bm{r}^A_\parallel \, e^{-i \bm{k}^R_\parallel \cdot \bm{r}^A_\parallel} D(\bm{r}^R_\parallel) \nonumber \\
    &= (2\pi)^2 \delta^{(2)}(\bm{k}^R_\parallel) D(\bm{r}^R_\parallel) \text{.} \label{eq:wigner_delta} 
\end{align}
\end{subequations}

The Wigner transform produces a delta function in the relative wavevector variable. 
We typically neglect both the delta function and the integral of $\bm{k}^R_\parallel$ in Eq.~\eqref{eq:wigner_delta}, 
which brings the GFs back to the cases described by Eq.~\eqref{eq:Fourier_definition} or Eq.~\eqref{eq:Fourier_definition_2}. 

\section{Derivation of Angular Momentum Transport Formula}

The derivation of the angular momentum flux density formula will be elaborated, starting from the Maxwell stress tensor expression $\langle yT_{xz}-xT_{yz}\rangle$. 
More precisely, the expression should be written as an integral, $\langle:\textstyle\int (yT_{xz}-xT_{yz}) \d x\d y:\rangle$. 
If we assume that the position coordinates are not quantum variables, we can use the linearity of the quantum average to simplify it to
$\textstyle\int (y\langle T_{xz}\rangle - x\langle T_{yz}\rangle) \d x\d y$. 
The expectations of $\langle T_{xz}\rangle$ and $\langle T_{yz}\rangle$ expressed in terms of the GFs are
\begin{subequations}
\begin{align}
    &\langle T_{xz}\rangle(z=z',\omega,\bm{k}_\parallel) \nonumber \\
    =&\frac{i\hbar}{\mu_0}(k_{y} k_{x} D^<_{yz} +i\partial_{z} k_{x} D^<_{yy}-k_{y} k_{y} D^<_{xz} -i\partial_{z} k_{y} D^<_{xy} \nonumber \\
    &+k_{x} k_{y} D^<_{zy} -k_{y} k_{y} D^<_{zx}-ik_{x} \partial_{z} D^<_{yy}+ik_{y}\partial_{z} D^<_{yx}  \nonumber\\
    &+\frac{\omega^2}{c^2} D^<_{zx}+\frac{\omega^2}{c^2} D^<_{xz}) \text{,} \label{eq:txz} \\
    &\langle T_{yz}\rangle(z=z',\omega,\bm{k}_\parallel)  \nonumber\\
    =& \frac{i\hbar}{\mu_0}(-i\partial_{z} k_{x} D^<_{yx} -k_{x} k_{x} D^<_{yz}+i\partial_{z} k_{y} D^<_{xx} +k_{x}k_{y} D^<_{xz} \nonumber\\
    &+ik_{x} \partial_{z} D^<_{xy} -ik_{y}\partial_{z} D^<_{xx}-k_{x} k_{x} D^<_{zy} +k_{y} k_{x} D^<_{zx} \nonumber\\
    &+\frac{\omega^2}{c^2} D^<_{zy}+\frac{\omega^2}{c^2} D^<_{yz}) \text{.} \label{eq:tyz}
\end{align}
\end{subequations}
However, $\langle T_{xz} \rangle$ and $\langle T_{yz} \rangle$ are independent of position, causing the integral to vanish because of the odd symmetry in the integrand.
We must therefore handle position coordinates as quantum variables and extract the non-zero terms cautiously.
Instead, we first evaluate the integrals, $\langle \textstyle\int yT_{xz} \d x\d y\rangle$ and $\langle \textstyle\int xT_{yz} \d x\d y\rangle$.
This expression should be cast in terms of the GF multiplied by the transverse coordinates, e.g., $x D^<(\bm{r},\bm{r}')$. 
Therefore, the integral of $x$ times the partial derivative of the GF can be expressed as
\begin{align}
    &\int \d x\d y \, x D^{<}_{\mu\nu}(\bm{r},t,\bm{r}',t') \nonumber\\
    =& \int \d x\d y \int \frac{\d^2 \bm{k}^R_\parallel}{(2\pi)^2} \int \frac{\d^2 \bm{k}^A_\parallel}{(2\pi)^2} \int \frac{\d\omega}{2\pi} \nonumber\\
    &\times e^{i(\bm{k}^A_\parallel+\bm{k}^R_\parallel/2)\cdot\bm{r}_\parallel} e^{-i(\bm{k}^A_\parallel-\bm{k}^R_\parallel/2)\cdot\bm{r}'_\parallel} e^{-i\omega (t-t')} \nonumber\\
    &\times x D^{<}_{\mu\nu}(z,z',\omega,\bm{k}^A_\parallel,\bm{k}^R_\parallel) \text{.} 
\end{align}

We also need to consider partial derivative operators.
For the Fourier transform, the partial derivative $\partial_{i'}$ and $\partial_j$ for the in-plane directions ($x,y$) are replaced by $-ik'_i$ and $ik_j$ respectively. 
For the $z$ direction, the corresponding rules are $\partial_{z'} \to -i \gamma' \sgn(z)$ and $\partial_{z} \to i \gamma \sgn(z)$,
\begin{equation}
\begin{split}
    &\int \d x\d y \, x \partial_{i'}\partial_j D^{<}_{\mu\nu}(\bm{r},t,\bm{r}',t') \\
    =& \int \d x\d y \int \frac{\d^2 \bm{k}^R_\parallel}{(2\pi)^2} \int \frac{\d^2 \bm{k}^A_\parallel}{(2\pi)^2} \int \frac{\d\omega}{2\pi} \\
    &\times e^{i\bm{k}^A_\parallel\cdot \bm{r}^R_\parallel} e^{i\bm{k}^R_\parallel\cdot\bm{r}^A_\parallel} e^{-i\omega (t-t')} \\
    &\times k_i' k_j x D^{<}_{\mu\nu}(z,z',\omega,\bm{k}^A_\parallel,\bm{k}^R_\parallel) \text{.}
\end{split}
\end{equation}
We substitute the variable $x$ with $x^A$ because of the condition that $\bm{r}'_\parallel\to \bm{r}_\parallel$.
Furthermore, the $x^A$ factor can be extracted by differentiating the exponential, $x^A\,e^{i\bm{k}^R_\parallel\cdot\bm{r}^A_\parallel}=-i\frac{\partial}{\partial k^R_x}e^{i\bm{k}^R_\parallel\cdot\bm{r}^A_\parallel}$. 
So, working backwards, we have 
\begin{align}
    &\int \d x\d y \, x \partial_{i'}\partial_j D^{<}_{\mu\nu}(\bm{r},t,\bm{r}',t') \nonumber\\
    =& \int \d x\d y \int \frac{\d^2 \bm{k}^R_\parallel}{(2\pi)^2} \int \frac{\d^2 \bm{k}^A_\parallel}{(2\pi)^2} \int \frac{\d\omega}{2\pi} \nonumber\\
    &\times 
    \,e^{i\bm{k}^A_\parallel\cdot \bm{r}^R_\parallel} (-i\frac{\partial}{\partial k^R_x}e^{i\bm{k}^R_\parallel\cdot\bm{r}^A_\parallel}) e^{-i\omega (t-t')} \nonumber\\ 
    &\times k'_i k_j D^{<}_{\mu\nu}(z,z',\omega,\bm{k}^A_\parallel,\bm{k}^R_\parallel) \text{.}
\end{align}
For the translationally invariant GF, the important relation is Eq.~\eqref{eq:wigner_delta}, in which the delta function can be expressed in the position integral form. We denote the new position coordinate as $\bm{R}_\parallel$, which is independent of $x$ and $y$, that is $(2\pi)^2 \delta^{(2)}(\bm{k}^R_\parallel)= \textstyle\int \d^2 \bm{R}_\parallel \, e^{-i \bm{k}^R_\parallel \cdot \bm{R}_\parallel}$.
We are interested solely in the lesser GFs in the limit $\bm{r}'_\parallel\to \bm{r}_\parallel$. In other words, this is equivalent to setting $\bm{r}^R_\parallel = 0$ and $\bm{r}^A_\parallel = \bm{r}'_\parallel = \bm{r}_\parallel$.
Under these conditions, the integral involving $x$ and $y$ becomes
$\textstyle\int \d x\d y \, e^{i\bm{k}^A_\parallel \cdot \bm{r}^R_\parallel} \frac{\partial}{\partial k^R_x}e^{i\bm{k}^R_\parallel\cdot\bm{r}^A_\parallel} = (2\pi)^2 \frac{\partial}{\partial k^R_x} \delta^{(2)}(\bm{k}^R_\parallel)$, i.e.,
\begin{align}
    &\int \d^2 \bm{R}_\parallel \int \frac{\d^2 \bm{k}^A_\parallel}{(2\pi)^2} \int \frac{\d^2 \bm{k}^R_\parallel}{(2\pi)^2} \int \frac{\d\omega}{2\pi} \nonumber\\
    &\times e^{-i\bm{k}^R_\parallel \cdot \bm{R}_\parallel} e^{-i\omega (t-t')} k'_i k_j D^{<}_{\mu\nu}(z-z',\omega,\bm{k}^A_\parallel) \nonumber\\
    &\times \int \d x \d y \, e^{i\bm{k}^A_\parallel\cdot \bm{r}^R_\parallel} (-i\frac{\partial}{\partial k^R_x} e^{i\bm{k}^R_\parallel\cdot\bm{r}^A_\parallel}) \nonumber\\
    =& -\int \d^2 \bm{R}_\parallel \int \d^2 \bm{k}^R_\parallel \int \frac{\d^2 \bm{k}^A_\parallel}{(2\pi)^2}  \int \frac{\d\omega}{2\pi} \nonumber\\
    &\times e^{-i \bm{k}^R_\parallel\cdot \bm{R}_\parallel} e^{-i\omega (t-t')} k'_i k_j D^{<}_{\mu\nu}(z-z',\omega,\bm{k}^A_\parallel) \nonumber\\
    &\times i \frac{\partial}{\partial k^R_x} \delta^{(2)}(\bm{k}^R_\parallel) \text{.}
\end{align}
In calculating the integral of $\bm{k}^R_\parallel$, because of the delta function, the integral will be $\textstyle\int f(\bm{k}^R_\parallel)\delta(\bm{k}^R_\parallel)\d \bm{k}^R_\parallel=f(0)$ and $\textstyle\int f(\bm{k}^R_\parallel)\delta'(\bm{k}^R_\parallel)\d \bm{k}^R_\parallel=-f'(0)$.
Thus, the partial derivatives of $k^R_x$ split the formula into two terms:
\begin{align}
        &\int\d^2 \bm{k}^R_\parallel \int\frac{\d^2 \bm{k}^A_\parallel}{(2\pi)^2} \int\frac{\d\omega}{2\pi}(\int \d\bm{R}^2_\parallel i\frac{\partial}{\partial k^R_x}e^{-i\bm{k}^R_\parallel\cdot \bm{R}_\parallel}) \nonumber\\
        &\times k'_i k_j D^{<}_{\mu\nu}(z-z',\omega,\bm{k}^A_\parallel)\delta^{(2)}(\bm{k}^R_\parallel)\Big|_{\bm{k}^R_\parallel=0} \nonumber\\
        &+\int \d^2 \bm{k}^R_\parallel \int\frac{\d^2 \bm{k}^A_\parallel}{(2\pi)^2} \int\frac{\d\omega}{2\pi}(\int \d^2 \bm{R}_\parallel e^{-i\bm{k}^R_\parallel\cdot \bm{R}_\parallel}) \nonumber\\ 
        &\times  i\frac{\partial}{\partial k^R_x}(k'_i k_j) \delta^{(2)}(\bm{k}^R_\parallel) D^{<}_{\mu\nu}(z-z',\omega,\bm{k}^A_\parallel)\Big|_{\bm{k}^R_\parallel=0} \nonumber\\
        =&\int R_{x} \d^2 \bm{R}_\parallel \int \frac{\d^2 \bm{k}_\parallel}{(2\pi)^2} \int\frac{\d\omega}{2\pi} k_i k_j D^{<}_{\mu\nu}(z-z',\omega,\bm{k}_\parallel) \nonumber\\
        &+\int \d^2 \bm{R}_\parallel \int \frac{\d^2 \bm{k}_\parallel}{(2\pi)^2} \int\frac{\d\omega}{2\pi} i\frac{\partial}{\partial k^R_x}(k'_i k_j)\Big|_{\bm{k}^R_\parallel=0} \nonumber\\
        &\times D^{<}_{\mu\nu}(z-z',\omega,\bm{k}_\parallel).
\end{align}
The first term corresponds to the case $\textstyle\int x\langle T_{yz}\rangle \d x\d y$, which has been previously discussed and yields $0$.
Consequently, only the second term is relevant, which represents the total angular momentum.
Meanwhile, when we set $\bm{k}^R_\parallel=0$, we also set $\bm{k}^A_\parallel=\bm{k}'_\parallel=\bm{k}_\parallel$. It turns the GF representation back into the form of the translationally invariant function.
Furthermore, we interpret the integral of $\bm{R}_\parallel$ as the product of flux density and area $\Sigma$~\cite{zym}
\begin{align}
        &\int \d x\d y\, x \partial_{i'}\partial_j D^{<}_{\mu\nu}(\bm{r},t,\bm{r}',t') \nonumber\\
        =& \Sigma \times \int \frac{\d^2 \bm{k}_\parallel}{(2\pi)^2} \int \frac{\d\omega}{2\pi} i\frac{\partial}{\partial k^R_x}(k'_i k_j)\Big|_{\bm{k}^R_\parallel=0} \nonumber\\
        &\times D^{<}_{\mu\nu}(z-z',\omega,\bm{k}_\parallel) \text{.} 
\end{align}
The calculation of the $y$-term is similar. After recalculating using Eqs.~\eqref{eq:txz} and \eqref{eq:tyz}, and with the transformation $k_x=k_\parallel$ and $k_y=0$, the simplified formula for $\langle yT_{xz}-xT_{yz} \rangle$ is 
\begin{align}
        &\langle yT_{xz}-xT_{yz} \rangle \nonumber\\
        =&\frac{\hbar}{\mu_0}
        (\frac{\partial}{\partial k^R_x}\gamma'k_x\sgn(z)D^<_{yx}+\gamma'\frac{\partial}{\partial k^R_x}k_x\sgn(z)D^<_{yx} \nonumber\\
        &+\frac{\partial}{\partial k^R_x}k'_{x} \gamma \sgn(z)D^<_{xy}+k'_{x} \frac{\partial}{\partial k^R_x}\gamma \sgn(z)D^<_{xy} \nonumber\\
        &-\frac{\partial}{\partial k^R_y} k'_{y} k_x D^<_{yz}-\gamma' \frac{\partial}{\partial k^R_y}k_{y}\sgn(z) D^<_{xy} \nonumber\\
        &-k'_{x}\frac{\partial}{\partial k^R_y}k_{y}D^<_{zy}-k'_{y} \frac{\partial}{\partial k^R_y}\gamma \sgn(z)D^<_{yx})\Big|_{\bm{k}^R_\parallel=0} \nonumber\\
        =&\frac{\hbar}{\mu_0}
        (\frac{k_\parallel}{2\gamma}k_\parallel\sgn(z)D^<_{yx}+\frac{1}{2}\gamma\sgn(z) D^<_{yx}-\frac{1}{2}\gamma\sgn(z) D^<_{xy} \nonumber\\
        &-k_\parallel\frac{k_\parallel}{2\gamma}\sgn(z)D^<_{xy}+\frac{k_\parallel}{2} D^<_{yz}-\gamma\frac{1}{2}\sgn(z)D^<_{xy} \nonumber\\
        &-\frac{k_\parallel}{2}D^<_{zy}+\frac{1}{2}\gamma\sgn(z)D^<_{yx}).
\end{align}
Finally, considering Eq.~\eqref{eq:z_component_dless}, we have
\begin{equation}
    \langle yT_{xz}-xT_{yz} \rangle =-\frac{\hbar}{\mu_0}\gamma\sgn(z)(D^<_{xy}-D^<_{yx}).
\end{equation}

\bibliography{references}

\end{document}